\documentclass[aps,superscriptaddress,twocolumn]{revtex4}

\usepackage[pdftex]{graphicx}
\usepackage{dcolumn}
\usepackage{bm}
\usepackage{color}
\usepackage{amsmath}
\usepackage{nicefrac}
\usepackage{amssymb} 

\newcommand{\ff}[1]{{\boldsymbol #1}}

\newcommand{\bi}{\begin{itemize}}
\newcommand{\ei}{\end{itemize}}
\newcommand{\be}{\begin{equation}}
\newcommand{\ee}{\end{equation}}
\newcommand{\ba}{\begin{eqnarray}}
\newcommand{\ea}{\end{eqnarray}}

\newcommand{\gap}{\Delta}

\usepackage[pdftex,colorlinks=true,linkcolor=blue,citecolor=blue,filecolor=blue]{hyperref}

\begin{document} 
  
\title{Long-time relaxation dynamics of a spin coupled to a Chern insulator}

\author{Michael Elbracht} 

\affiliation{I. Institute of Theoretical Physics, Department of Physics, University of Hamburg, Jungiusstra\ss{}e 9, 20355 Hamburg, Germany}

\author{Michael Potthoff}

\affiliation{I. Institute of Theoretical Physics, Department of Physics, University of Hamburg, Jungiusstra\ss{}e 9, 20355 Hamburg, Germany}

\affiliation{The Hamburg Centre for Ultrafast Imaging, Luruper Chaussee 149, 22761 Hamburg, Germany}

\begin{abstract}
The relaxation of a classical spin, exchange coupled to the local magnetic moment at an edge site of the one-dimensional spinful Su-Schrieffer-Heeger model is studied numerically by solving the full set of equations of motion. 
A Lindblad coupling of a few sites at the opposite edge to an absorbing bath ensures that convergence with respect to the system size is achieved with only a moderate number of core sites. 
This allows us to numerically exactly study the long-time limit and to determine the parameter regimes where spin relaxation takes place. 
Corresponding dynamical phase diagrams for the topologically trivial and the nontrivial cases are constructed. 
The dynamical phase boundaries, the role of the topological edge state and its internal Zeeman splitting for the spin-relaxation process, as well as incomplete spin relaxation on long time scales can be explained within the framework of a renormalized linear-response approach when explicitly taking retardation effects and nonequilibrium spin-exchange processes into account. 
\end{abstract} 

\maketitle 

\section{Introduction}
\label{sec:intro}

Novel concepts \cite{NYP+17} to achieve ever smaller magnetic bits and thus higher data storage continue to drive research of  systems of magnetic atoms on nonmagnetic surfaces \cite{Wie09}.
Since the manipulation of magnetic bits requires external time-dependent fields, there is a strong 
interest in the stability of excitations of single magnetic atoms. 
Such spin excitations of single absorbed magnetic atoms can be probed experimentally, e.g., via inelastic scanning tunneling spectroscopy \cite{HGLE04,HLH06,Fra09,Fer09,GLN12}.

Surfaces of topological insulators \cite{HK10,QZ11} are particularly interesting in this context since a magnetic impurity atom located at the surface is expected to predominantly interact with the conducting surface state and since the existence of this surface state and its robustness against weak perturbations is ensured by the topological properties of the bulk band structure and the bulk-boundary correspondence principle \cite{FJK11,MS11,FSFF12,PSB16}.
The {\em static} properties of magnetic impurities at the surface of topological insulators have been studied extensively, both experimentally and theoretically \cite{WXX11,HKS+12,SSM+12,VPG+12,GLA13,SBK+13,EWS+14,LZLZ14,JSL+15,CTH+15,PF16,RMS+18,SKR+19}.

Recently, also {\em dynamical} properties of impurities at surfaces of topological insulators have been investigated theoretically, based on the linear-response approach within time-dependent density-functional theory \cite{BDGL19} and on Floquet theory applied to a periodically driven (nonmagnetic) impurity coupled to a two-dimensional topological insulator \cite{PF19}.
Earlier theoretical studies have considered the effect of the surface state of a topological insulator on the  magnetization dynamics of a coupled ferromagnetic system \cite{GF10,YZN10,UTTY12,TL12}. 
An array of magnetic adatoms interacting with the electronic surface states was investigated in Ref.\ \cite{CED+14}.
A large single-atom anisotropic magnetoresistance on a surface of a three-dimensional topological insulator (Bi$_{2}$Se$_{3}$) decorated with magnetic adatoms (Mn) has been found in first-principles transport calculations \cite{NRS15}.
Beyond the level of linear-response theory, however, the full microscopic real-time dynamics of a single magnetic atom coupled to the electronic structure of the topological substrate has not been addressed so far. 

Clearly, for real systems, the application of non-perturbative time-dependent {\em ab initio} methods is extremely demanding.
The situation is different, however, in case of strongly simplified model systems, where one can address the full real-time dynamics of an initial magnetic excitation beyond the linear-response approach.
Here, we consider the one-dimensional spinful Su-Schrieffer-Heeger (SSH) model \cite{SSH80,HKSS88,AOP16} as a prototypical system, which, depending on the ratio of the hopping parameters, hosts a Kramers-degenerate edge state at each of the boundaries.
Coupling a classical spin to one of the edge sites locally destroys time-reversal symmetry, which leads to a spin splitting of the edge state. 
There is a closed system of equations of motions such that, in principle, the full coupled real-time dynamics 
\cite{AKvSH95,KFN09,PSS11,SP15,CBW+18,BN19} of  the electronic structure and the classical spin is accessible by numerical means beyond the linear-response theory \cite{ON06,BNF12,UMS12}.

However, even in this comparatively simple case, calculations based on the full set of equations of motion are demanding since relaxation times typically exceed the bare electronic time scale by several orders of magnitudes \cite{AKvSH95,SP15}.
We note that estimates for lifetimes of excitations of 3d and 4d magnetic impurities embedded in Bi$_{2}$Te$_{3}$ and Bi$_{2}$Se$_{3}$ range from the pico- to the microsecond regime \cite{BDGL19}. 
Let us also mention that simulations of real-time dynamics based on classical {\em spin-only models}, see Refs.\ \cite{SHNE08,EFC+14} for example, are much simpler and can be performed for large two- or three-dimensional systems approaching the thermodynamic limit \cite{MDSW10} or coupling the spin system to classical lattice degrees of freedom in addition \cite{LW19}.

Here, on the other hand, we are interested in the dynamic relaxation process of a classical spin coupled to a topologically nontrivial electronic structure.
The necessity to explicitly account for the time dependence of the electronic structure complicates the computations. 
Due to spin and energy conservation, the relaxation of an initial magnetic excitation requires the transport of spin and energy away from the magnetic impurity and dissipation into the bulk of the system. 
An exact treatment of the equations of motion, however, can only be done for a finite, comparatively small system size in practice.
This implies that excitations of the electronic system that are emitted by the impurity will eventually reach the boundaries of the system.
Reflections at the boundaries, back-propagation and interference with the dynamics close to the impurity will severely spoil the computation of the spin-relaxation time. 

Recently, we have constructed \cite{EP20} a novel type of absorbing boundary conditions, which employs a generalized Lindblad master-equation approach to couple the edge sites of the conduction-electron tight-binding model to an external bath.
With these boundary conditions, outgoing excitations resulting from an initial excitation of a classical spin exchange-coupled to the conduction-electron system can be absorbed completely without disturbing the dynamics close to the impurity spin. 
It has been demonstrated that this allows us to trace the spin and the conduction-electron dynamics on a time scale, which exceeds the characteristic electronic scale that is set by the inverse nearest-neighbor hopping by more than five orders of magnitude.

Here, we will employ these absorbing boundary conditions to microscopically trace the coupled time evolution of spin and electron degrees of freedom for a single classical spin coupled to one of the edges of an SSH model on long time scales.
In particular, we study the impact of the electronic edge state on the spin relaxation time.
Lowest-order time-dependent perturbation theory in the exchange coupling is expected to break down in the long-time regime. 
We therefore carefully check the validity of the linear-response approach and head for possible new non-perturbative phenomena.

\section{Classical spin coupled to the spinful SSH model}
\label{sec:mod}

\begin{figure}[t]
\includegraphics[width=0.75\columnwidth]{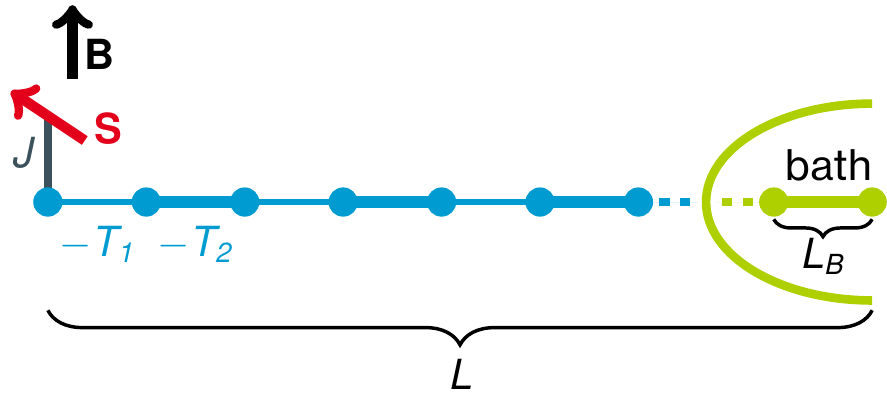}
\caption{
Sketch of the system geometry:  
A classical spin $\bm{S}$ of length $|\bm{S}| = \frac{1}{2}$ is coupled via a local antiferromagnetic
exchange interaction $J$ to the first site of a spinful SSH model consisting of $L$ sites.
The nearest-neighbor hopping alternates between $-T_1$ and $-T_2$.
To construct absorbing boundary conditions at the opposite edge, the last $L_B$ sites of the chain are coupled to a Lindblad bath \cite{EP20}.
The classical spin is subjected to a local magnetic field $\bm{B}$. 
At time $t=0$, real-time dynamics is initiated by a sudden flip of the field direction.
}
\label{fig:sys}
\end{figure}

Fig.\ \ref{fig:sys} presents a sketch of the setup considered here. 
The corresponding Hamiltonian is given by $H=H_{0} + H_{\rm imp}$, where
\be 
  H_{0} = - \sum_{i=1}^{L-1} \sum_{\sigma=\uparrow, \downarrow} 
  [(T + (-1)^{i} \delta T) c^{\dagger}_{i\sigma} c_{i+1\sigma} + \mbox{H.c.}]
\label{eq:h0}
\ee
is the spinful SSH model \cite{SSH80,HKSS88,AOP16}. 
An electron at site $i$ with spin projection $\sigma=\uparrow, \downarrow$ is created or annihilated by $c^{\dagger}_{i\sigma}$ or $c_{i\sigma}$, respectively.
The nearest-neighbor hopping amplitudes alternate between 
\begin{align}
\begin{split}
 T_{1} \equiv T - \delta T \quad \mbox{and} \quad T_{2}\equiv T + \delta T \: .
\label{eq:t12}
\end{split}
\end{align}
We set $T=1$ to fix the energy unit and (with $\hbar \equiv 1$) the time unit. 
Furthermore, the chemical potential is set to $\mu = 0$ such that the system is at half-filling with $N=L$ electrons. 
For $\delta T = 0$, the system is in a metallic state, while for any nonzero $\delta T$ there is finite a bulk band gap 
\be
\gap = 4 \, |\delta T| = 2 \, |T_1-T_2|
\label{eq:gap}
\ee
between the highest occupied and the lowest unoccupied state. 
For even $L$ and for $\delta T < 0$, i.e.\ for $T_{1}>T_{2}$, the system is in a topologically trivial insulating state. 
If $\delta T > 0$, on the other hand, the system is topologically nontrivial, and two spin-degenerate edge states appear at zero energy. 
These are exponentially localized close to each of the two edges $i=1$ and $i=L$.
We will couple the sites close to the $i=L$ edge to an absorbing bath. 
When studying the topologically nontrivial case, we will therefore consider a chain with an odd number of sites $L$ and $\delta T>0$. 
In this case there is a spin-degenerate edge state localized close to the $i=1$ edge only. 
This is a convenient choice which does not affect the physics close to the $i=1$ edge.

The impurity part of the Hamiltonian is given by
\be 
  H_{\rm imp} = J \ff s_{i_{0}} \ff S - \ff B \ff S \: .
\label{eq:himp}
\ee
Here, $\ff S = (S_{x},S_{y},S_{z})$ denotes classical spin of length $S=\frac12$. 
This is coupled via a local antiferromagnetic ($J>0$) exchange to the local spin of the electron system $\ff s_{i_{0}} = \frac12 \sum_{\sigma\sigma'} c^{\dagger}_{i\sigma} \ff \tau_{\sigma\sigma'} c_{i\sigma'}$ at site $i_{0}$. 
Here, $\ff \tau$ is the vector of Pauli spin matrices. 
Throughout the study, we will couple the classical spin to the edge site $i_{0}=1$.
Furthermore, the model includes an external local magnetic field $\ff B$ which is used to drive the classical spin.

\section{Equations of motion and absorbing boundaries}
\label{sec:eom}

The motion of the classical spin is driven by the torque that is generated by the external local field $\ff B$ and by the torque due to the local magnetic moment $\langle \ff s_{i_{0}} \rangle = \langle \Psi(t) |\ff s_{i_{0}}|\Psi(t)$ of the electron system. 
Here, $|\Psi(t) \rangle$ is the $N$-electron state at time $t$. 
With the help of the one-particle reduced density matrix $\ff \rho(t)$ defined as 
\be
\rho_{i\sigma i' \sigma'}(t) = \langle \Psi(t) |c^{\dagger}_{i'\sigma'} c_{i\sigma}| \Psi(t) \rangle
\: , 
\label{eq:dens}
\ee
the equation of motion for the classical spin can be written as:
\be
  \frac{d}{dt} \bm{S}(t) = J \langle \bm{s}_{i_0} \rangle_t \times \bm{S}(t) - \bm{B} \times \bm{S}(t)
\: .
\label{eq:eoms}
\ee
The one-particle reduced density matrix satisfies a von Neumann-type equation of motion, 
\be
i \frac{d}{dt} \ff \rho(t) =  [\ff T^{(\rm eff)}(t), \ff \rho(t)]  \: , 
\label{eq:eomr}
\ee
where $\ff T^{(\rm eff)}(t)$ is an effective, time-dependent hopping matrix with the elements
\be
  T^{\rm (eff)}_{i\sigma i'\sigma'}(t) = T_{ii'} \delta_{\sigma\sigma'} + \delta_{ii'} \frac J2 \ff S(t) \ff \tau_{\sigma\sigma'} 
  \: , 
\label{eq:teff}
\ee
and where $T_{ii'}$ are the elements of the standard hopping matrix $\ff T$. 
Its nonzero elements are given by $T_{i,i+1}=-(T + (-1)^{i} \delta T) = T_{i+1,i}$ for $i=1,...,L-1$.
Let us emphasize that we do not have to construct the $N$-electron state $|\Psi(t) \rangle$ explicitly, since Eqs.\ (\ref{eq:eoms}) and (\ref{eq:eomr}) form a closed nonlinear set of ordinary differential equations. 
This is due to the fact that the electron system is effectively non-interacting such that $|\Psi(t)\rangle$ is a simple Slater determinant at any instand of time $t$.
In Refs.\ \onlinecite{Elz12,SP15} the foundations of the dynamics of quantum-classical hybrid systems and the concrete derivation of the equations of motions are discussed in detail.
We note that the equations of motion (\ref{eq:eoms}) and (\ref{eq:eomr}) imply the conservation of the total energy $\langle H \rangle$, the total particle number $N=\sum_{i\sigma} \langle c_{i\sigma}^{\dagger} c_{i\sigma} \rangle$ and (for B=0) the total spin $\ff S + \sum_{i} \langle \ff s_{i} \rangle$.

At time $t=0$ the system is prepared in the ground state for a field pointing in $x$-direction. 
This aligns the classical spin, $\ff S(t=0) \propto \ff e_{x}$. 
We set the according effective hopping matrix $\ff T^{\rm (eff)}(t=0)$ for the given spin direction, and, via numerical diagonalization of $\ff T^{\rm (eff)}(0)$, compute the initial density matrix as the corresponding ground-state density matrix: $\ff \rho(t=0) = \Theta(\mu \ff 1 - \ff T^{(\rm eff)}(0))$.
Here, $\Theta$ is the Heavyside step function and $\mu=0$. 
To initiate the dynamics, we then suddenly flip the field to $z$-direction and hold the field direction and strength constant for $t>0$.

We are interested in tracing the time evolution up to the point, where the system is fully relaxed, i.e., where locally, close to $i_{0}$, the system reaches its ground state. 
This defines the spin relaxation time $\tau$. 
As the computational effort for solving the equations of motion scales as $L^{3}$ for large $L$, one is in practice limited to a system size of $L \lesssim 1000$. 
For gapped systems discussed below, the relaxation time is typically large (up to $10^{5}$ inverse hoppings). 
This implies that one cannot avoid unwanted finite-size effects due to reflection of the excitations initiated at $i_{0}$ from the opposite system edge simply by taking a sufficiently large system. 

The problem can be solved, however, by using absorbing boundary conditions. 
As is discussed in detail in Ref.\ \cite{EP20}, these can be realized within the framework of the Lindblad master equation, i.e., by coupling the outmost $L_{B} \ll L$ sites $i = L - L_{B}+1, ..., L$ of the system to a bath that fully absorbs the spin and energy of any excitations emitted from the spatial region close to edge at $i=1$, where the classical spin is coupled to.
A naive application of the Lindblad equation, however, has been demonstrated as being inadequate. 
One must in fact carefully exclude artifacts that could be introduced at $t=0$ and early times due to the coupling to the bath itself. 
This can be taken care of with a matrix formulation of the Lindblad equation which still respects the Hermiticity and the nonnegativity of $\ff \rho(t)$ at all times $t$ and by a special choice of the (matrix) Lindblad parameters. 
Following our previous work \cite{EP20} this results in replacing Eq.\ (\ref{eq:eomr}) by 
\ba
  i \frac{d}{dt} \ff \rho(t) 
  =
  [ \ff T^{(\rm eff)}(t) , \ff \rho(t) ]
  - i
  \{\ff \gamma, \ff \rho(t) - \ff \rho(0) \}
  \: .
\label{eq:eomm}
\ea
The coupling of the outermost $L_{B}$ sites to the bath is regulated by the diagonal matrix $\ff \gamma$ with diagonal elements $\gamma_{i\sigma,i\sigma} = \gamma_{i}$ for $i = L - L_{B}+1, ..., L$ and 
$\gamma_{i\sigma,i\sigma}=0$ else. 
The precise site dependence of $\gamma_{i}$ is not very important. 
Here, we employ a linear profile: $\gamma_{i} = (i- (L-L_{B})) \gamma_{\rm min}$ for $i = L - L_{B}+1, ..., L$ with a single parameter $\gamma_{\rm min}$.
Clearly, when employing absorbing boundary conditions, energy and spin conservation only holds locally, for $i< L-L_{B}$. 
Due to manifest particle-hole symmetry, on the other hand, conservation of the total particle number still holds.

\section{Topologically trivial case}
\label{sec:tri}

\begin{figure}[t]
\centering
\includegraphics[width=0.8\columnwidth]{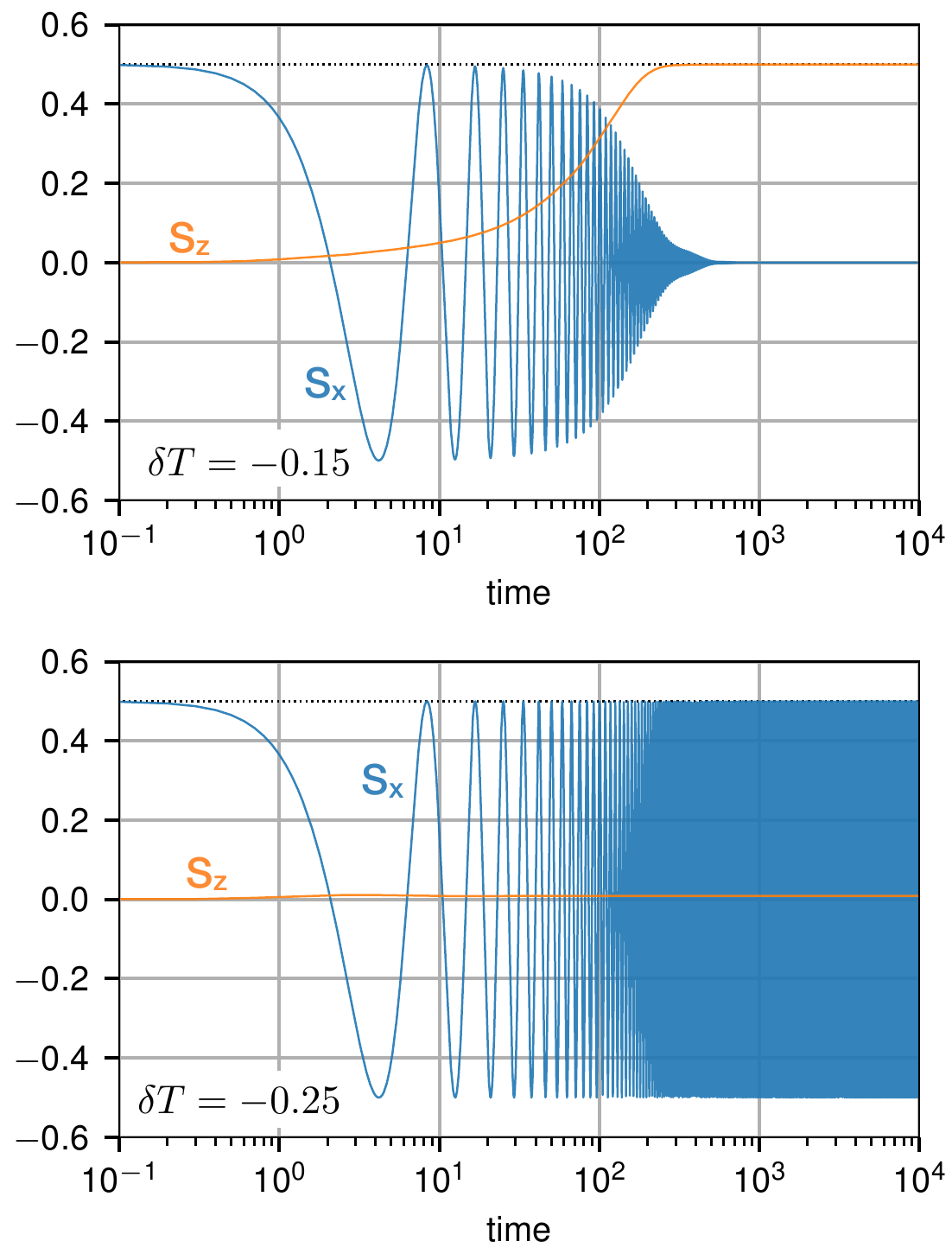}
\caption{
Real-time dynamics of the impurity spin after a sudden flip of the local magnetic field from $x$- to $z$-direction at time $t=0$. 
Calculations for a system with $L=46$ sites, exchange interaction $J=1$, field strength $B=0.75$. 
{\em Upper panel}: $\delta T=-0.15$, {\em lower panel:} $\delta T = -0.25$. 
Absorbing boundary conditions with $L_B=4$ and $\gamma_\text{min}=0.2$.
Energy and time units are set by the hopping integral $T\equiv 1$.
} 
\label{fig:relax}
\end{figure}

Numerical results will be discussed for the topologically trivial case $\delta T<0$ first. 
The upper panel of Fig.\ \ref{fig:relax} displays the time evolution of the $x$- and $z$-component of the  impurity spin for $\delta T = - 0.15$. 
After the sudden flip of the field $\ff B = B \ff e_{x} \mapsto B \ff e_{z}$, the spin immediately starts to precess around the $z$-axis, as can be seen in the $x$- and $y$-component (not shown) of $\ff S(t)$. 
The precession frequency is approximately given by the Larmor frequency $\omega_{\rm p} \approx B$.
Additionally, there is a damping effect due to the coupling of the spin to the electron system. 
This is seen in the $z$-component of $\ff S(t)$, which steadily increases with time until at $\tau \approx 500$ the system is fully relaxed and the spin is aligned to the new field direction. 
This is a plausible result which is qualitative similar to the spin dynamics seen in the case of a metallic electron system \cite{SP15,EP20}.

We have regularly checked the reliability of the calculations.
Results obtained with absorbing boundary conditions for system size $L$ are compared with those for much larger systems with open boundary conditions on the short time scale before reflections from the edge opposite to the impurity spin set in. 
We also compare calculations performed with absorbing boundary conditions for different system sizes and have checked that results do not depend on $L$.
Systems as small as $L=46$ turn out as fully sufficient for convergence.
The parameters $L_B$ and $\gamma_\text{min}$ are optimized to suppress backscattering of excitations from the opposite edge of the system. 
This is in fact the case for the parameter choices made here. 
A detailed discussion and is given in Ref.\ \cite{EP20}.

In the lower panel of Fig.\ \ref{fig:relax} the result of a calculation with the same parameters but for $\delta T=-0.25$ is shown. 
Here, we see the same precessional motion but there is almost no damping of the spin up a propagation time $t=10^{4}$. 
We conclude that the relaxation time crucially depends on the hopping parameters and thus on the size of the gap $\Delta = 4 |\delta T|$.

\begin{figure}[t]
\centering
\includegraphics[width=0.9\columnwidth]{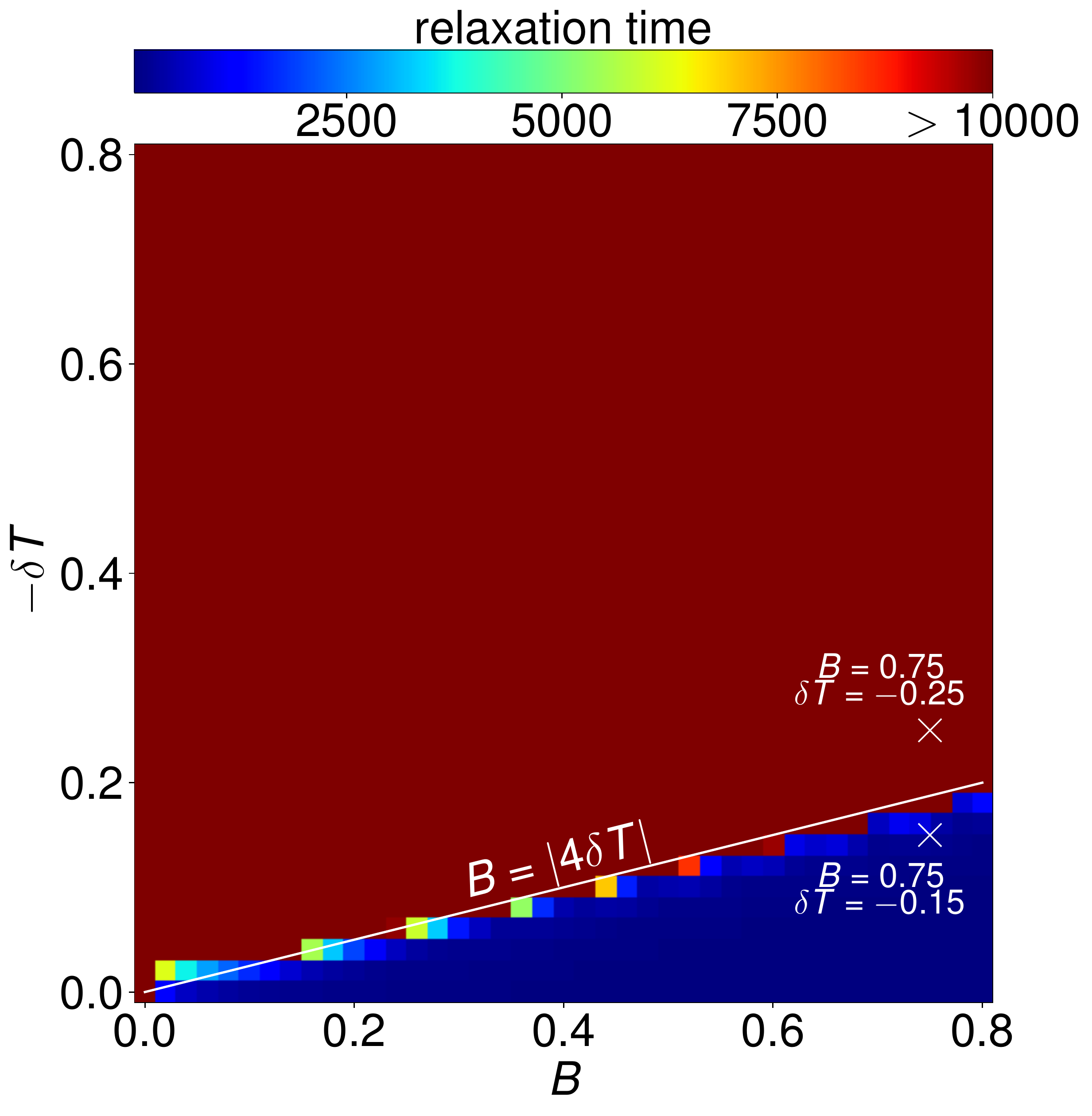}
\caption{
Relaxation time $\tau$ (see color code) as a function of $-\delta T$ and $B$. 
Note that the dark red region filling most of the phase diagram indicates relaxation times $\tau > 10^{4}$.
Other parameters as in Fig.\ \ref{fig:relax}.
White line: $B = 4|\delta T|$.
White crosses: parameter sets used in Fig.\ \ref{fig:relax}.
}
\label{fig:pd1}
\end{figure}

An overview is given with Fig.\ \ref{fig:pd1}. 
Here we display the spin relaxation time $\tau$ as a function of $-\delta T> 0$ and $B$.
Each point in the phase diagram is obtained from an independent calculation solving the full set of equations of motion (\ref{eq:eoms}) and (\ref{eq:eomr}).
For a definitive pragmatic definition, we fix $\tau$ as the shortest time for which the $z$-component of $\ff S$ is larger than 95\% of its fully relaxed value, i.e., $S_{z}(t>\tau) > 0.475$.
Other choices would not significantly affect the phase diagram and the interpretation. 
There is a relatively clear separation between parameter sets for which spin relaxation is seen and those where the spin is not relaxed on the maximum propagation time scale $t=10^{4}$ considered. 
For fixed $\delta T$ and with decreasing field strength $B$, the spin-relaxation time diverges at a critical field. 
This is quite precisely given by $B=4|\delta T|$, i.e., we find that the system relaxes, if $B > \gap = 4|\delta T|$.
We have also checked that this result does not depend on the initial direction of the classical spin. 
Calculations with initial directions $\bm{S}(t=0) = \frac{1}{2}(\cos \varphi,0, \sin \varphi)$ for various angles $\varphi$ yield essentially the same phase diagrams and, in particular, the same phase boundary separating parameter regions leading to full spin relaxation or not.

To understand the phase boundary, we first consider the excitation energy, which is pumped into the system at $t=0$ due to the sudden field switch. 
This is given by $E_{\rm ex} = E_{\rm fin} - E_{\rm ini}$, where 
$E_{\rm ini}$ is the ground-state energy of the whole system, and where $E_{\rm fin}$ is the energy right after the sudden flip of the field direction. 
Note that, due to total energy conservation, $E_{\rm fin} = \mbox{const.}$ for early times until excitations of the electron system are absorbed at the opposite boundary. 
Right after the switch of the field, the energy of the Fermi sea $\langle H_{0} \rangle$ and the exchange-coupling energy $J \langle \ff s_{i_{0}} \rangle \ff S$ are unchanged. 
Hence, $E_{\rm ex} = - \ff B_{\rm fin} \ff S - (- \ff B_{\rm ini} \ff S) = \ff B_{\rm ini} \ff S = B/2$, since $\ff B_{\rm fin} \perp \ff S$ and $|\ff S|=1/2$.
This energy must be dissipated to the bulk to achieve complete spin relaxation. 

Since the lowest excitation energy of the electron system is given by the gap $\Delta = 4 |\delta T|$, it is tempting to assume that spin relaxation is possible if $E_{\rm ex} > \Delta$, i.e., if $B > 8 |\delta T|$. 
Obviously, however, this argument cannot explain the different spin dynamics seen in the upper and the lower panel of Fig.\ \ref{fig:relax} -- it would predict absence of spin relaxation for both cases, $\delta T = -0.15$ and $\delta T = -0.25$ at $B=0.75$.
It is at variance with the phase boundary displayed in Fig.\ \ref{fig:pd1} by a factor of two.
In fact, the argument misses that the exchange coupling $J\ff s_{i_{0}}\ff S$ mediates virtual processes at order $J^{2}$.
Such processes can slightly tilt the classical spin and reduce its potential energy in the external field by an arbitrarily small amount.

This can be formalized by time-dependent perturbation or linear-response theory, see Refs.\ \cite{ON06,BNF12,UMS12,SP15,SRP16a}, for example:
In first order in $J$, the response of the local magnetic moment $\langle \ff s_{i_{0}} \rangle_{t}$ at time $t$ due to the classical spin $\ff S(t)$ is given via the Kubo formula as
\be
  \langle \ff s_{i_0} \rangle_t = J \int_0^t dt' \chi_{\rm loc}(t-t') \bm{S}(t') \; , 
\label{eq:lr}
\ee
where $\chi_{\rm loc}(t) \equiv \chi_{{\rm loc}, \alpha\alpha}(t)$ ($\alpha=x,y,z$) is the retarded isotropic local magnetic susceptibility
$\chi_{{\rm loc},\alpha \beta}(t-t') = - \Theta(t-t') \langle [ s_{i_0}^\alpha (t) , s_{i_0}^\beta (t') ] \rangle$.
Inserting Eq.\ (\ref{eq:lr}) in the equation of motion (\ref{eq:eoms}) for $\ff S(t)$, yields an integro-differential equation
\be
  \frac{d}{dt} \bm{S}(t) 
  = 
  \bm{S}(t) \times \bm{B} - J^2 \bm{S}(t) \times \int_0^t dt' \chi_{\rm loc}(t-t') \bm{S}(t')
  \: .
\label{eq:ide}  
\ee
This shows that spin damping originates from the second term as a memory effect. 
Furthermore, with Eq.\ (\ref{eq:ide}) it is straightforward to see that, disregarding a transient effect at early $t$, a precessional motion of the $x,y$-components of $\ff S$ with frequency $\omega_{\rm p} = B$ induces a change of its $z$-component $\dot{S}_{z} \propto J^{2} \mbox{Im} \, \chi_{\rm loc}(\omega=B)$, where $\chi_{\rm loc}(\omega) = \int d\tau \chi_{\rm loc}(\tau) e^{i\omega \tau} / 2 \pi$ is the frequency-dependent susceptibility.
This implies that spin damping is obtained if 
\be
  \mbox{Im} \, \chi_{\rm loc}(\omega=B) \ne 0 \: .
\label{eq:cond}  
\ee

\begin{figure}[t]
\centering
\includegraphics[width=0.75\columnwidth]{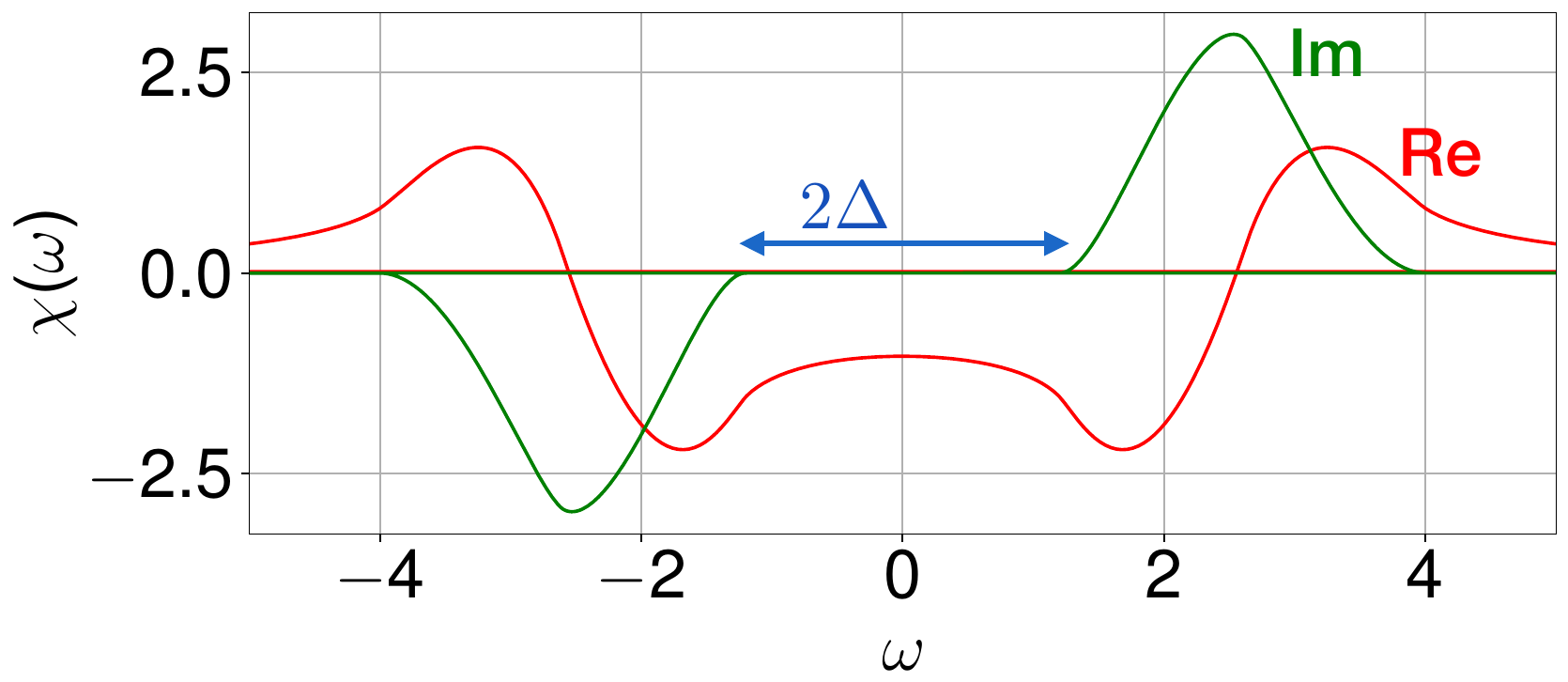}
\caption{
Imaginary and real part of the local retarded spin susceptibility $\chi_{\rm loc}$ as function of the excitation energy $\omega$ at the edge site $i=1$ of the SSH model.
Calculation for topologically trivial case $\delta T = -0.3$. 
}
\label{fig:chi}
\end{figure}

For a non-interacting system of electrons, the susceptibility $\chi_{\rm loc}(\omega)$ is given by a convolution of the occupied with the unoccupied part of the local density of states $\rho_{\rm loc}(\omega)$ (LDOS). 
Hence, one finds that spin excitations are gapped and that the spin gap is twice the one-particle excitation gap.
This can also be demonstrated explicitly by computing (see Ref.\ \cite{SP15})
\ba
\chi_{\rm loc}(t) 
&=&
\Theta(t) \mbox{Im} \Big[
\big(e^{-i\ff T t} \Theta(\ff T - \mu)\big)_{i_{0}i_{0}} 
\nonumber \\
&\cdot&
\big(e^{i\ff Tt} \Theta(\mu - \ff T)\big)_{i_{0}i_{0}}
\Big]
\: .
\ea
via numerical diagonalization of the unperturbed ($J=0$) hopping matrix $\ff T$ of the SSH model.
The Fourier transform is easily obtained numerically and shown in Fig.\ \ref{fig:chi}.
We see that the imaginary part vanishes for frequencies with $-\Delta < \omega  < \Delta$, where $\Delta = 4 |\delta T|$ is the one-particle excitation gap.
Therewith, the condition (\ref{eq:cond}) for spin damping reads as $B> \Delta = 4 |\delta T|$.
This nicely fits with the boundary in the dynamical phase diagram, see the white line in Fig.\ \ref{fig:pd1}.
We conclude that linear-response theory well describes the topologically trivial case.

\section{Topologically nontrivial case}
\label{sec:non}

When the spin is coupled to a site $i_{0} \approx L/2$ in the bulk of the system, its relaxation behavior is fully determined by the bulk electronic structure and by the bulk band gap in particular. 
This means that there is no difference in the spin dynamics upon a sign change of $\delta T$, cf.\ Eq.\ (\ref{eq:gap}).
This has been checked and verified numerically. 

However, in the topologically nontrivial case for $\delta T > 0$, there is a protected edge state localized around $i=1$. 
Its spin degeneracy is lifted due to the exchange coupling to the classical spin, which is coupled to the electron system at site $i_{0}=1$, and in the ground state at half-filling ($\mu=0$) the state of the spin doublet with lower eigenenergy is fully occupied. 
The presence of this polarized edge state is expected to affect the mechanism for the relaxation of the classical impurity spin.
We consider systems with an odd number of sites $L$ such that there is no edge state at the opposite edge, where a few sites are coupled to the absorbing bath. 

\subsection{Pre-relaxation}
\label{sec:pre}

\begin{figure}[t]
\centering
\includegraphics[width=0.9\columnwidth]{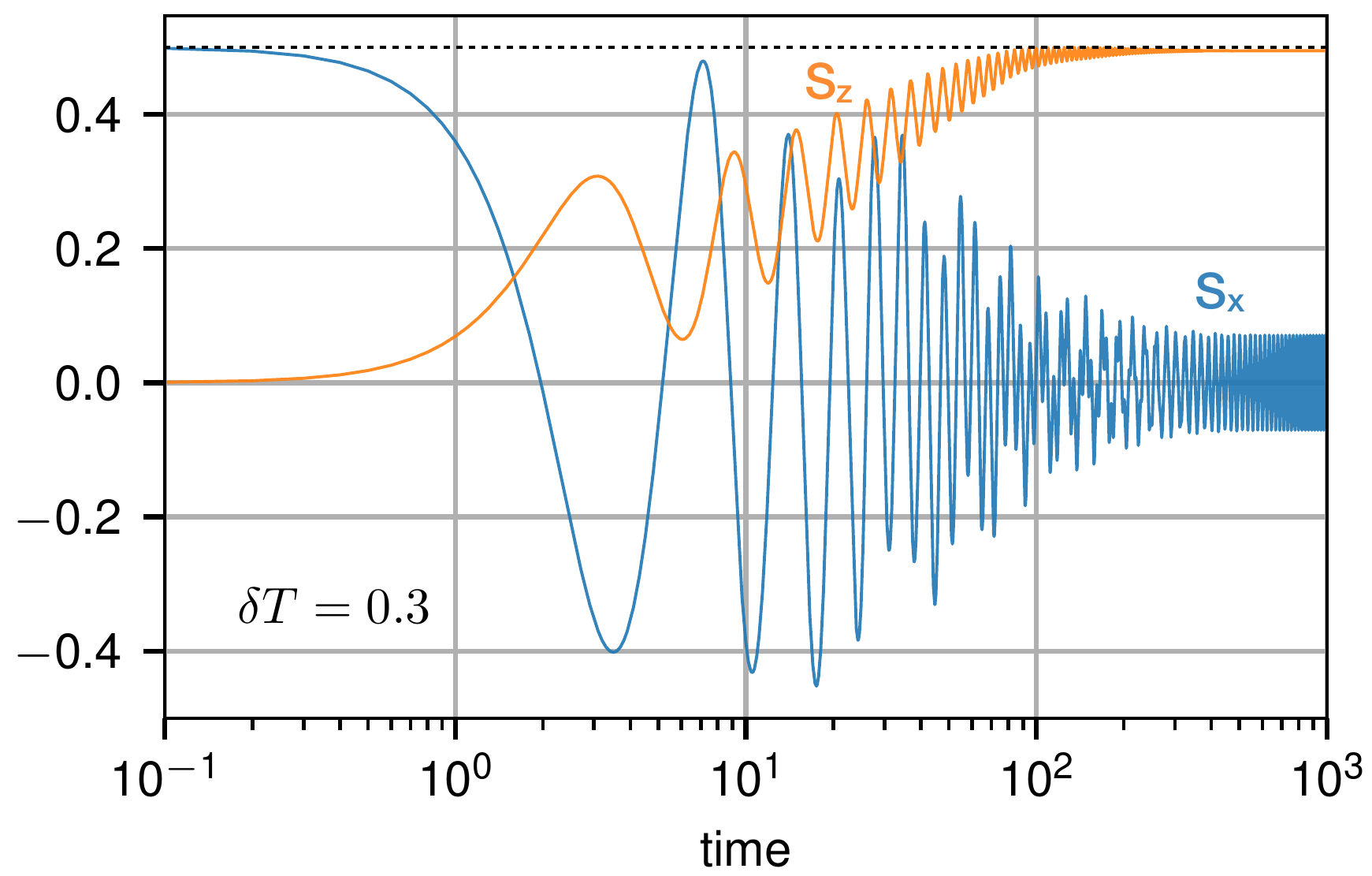}
\caption{
Impurity spin dynamics for $\delta T = 0.3 >0$ after a sudden flip of the local magnetic field from $x$- to $z$-direction at time $t=0$. 
Calculations for $L=47$, $J=1$, $B=0.75$. 
Absorbing boundary conditions with $L_B=5$ and $\gamma_\text{min}=0.2$.
} 
\label{fig:sd}
\end{figure}

An example for the impurity-spin dynamics is given with Fig.\ \ref{fig:sd}.
Compared to the topologically trivial case, see Fig.\ \ref{fig:relax}, the result is qualitatively different.
Overall, the time evolution is somewhat more complicated and composed of at least two oscillations with different frequencies, rather than one major oscillation as in the absence of the edge state.

At early times $t \lesssim 100$, a simple precessional dynamics is seen with only slight irregularities and with a well defined precession frequency which is close to but somewhat higher than the Larmor frequency $B$.
At later times the spin starts to relax and to align to the new field direction $\ff B = B \ff e_{z}$. 
In a temporal transition regime around $t=100$, the dynamics is less simple.

For $t\gg 100$, however, a quite regular dynamics is seen again. 
Here, $S_{z}$ approaches as constant. 
We find $S_{z}/S > 95\%$, i.e., spin relaxation according to our pragmatic definition above, but clearly $S_{z}$ stays smaller than its fully relaxed value $S_{z}=S$. 
In fact, up to a time scale of $\tau = 1 \cdot 10^{5}$, we do not see any indication for a complete relaxation of the spin. 
$S_{z}$ rather saturates at a value $S_{z} \approx 0.495<S$.
We refer to this behavior as ``pre-relaxation''. 
Accordingly, the transversal $x$- and $y$-components undergo a precessional motion. 
The corresponding frequency is $\omega_{\rm p} \approx 0.285$, i.e., much smaller than the Larmor frequency $\omega_{\rm p} = B = 0.75$. 
The cause of the incomplete relaxation is discussed below in Sec.\ \ref{sec:inc}.

\begin{figure}[t]
\centering
\includegraphics[width=0.9\columnwidth]{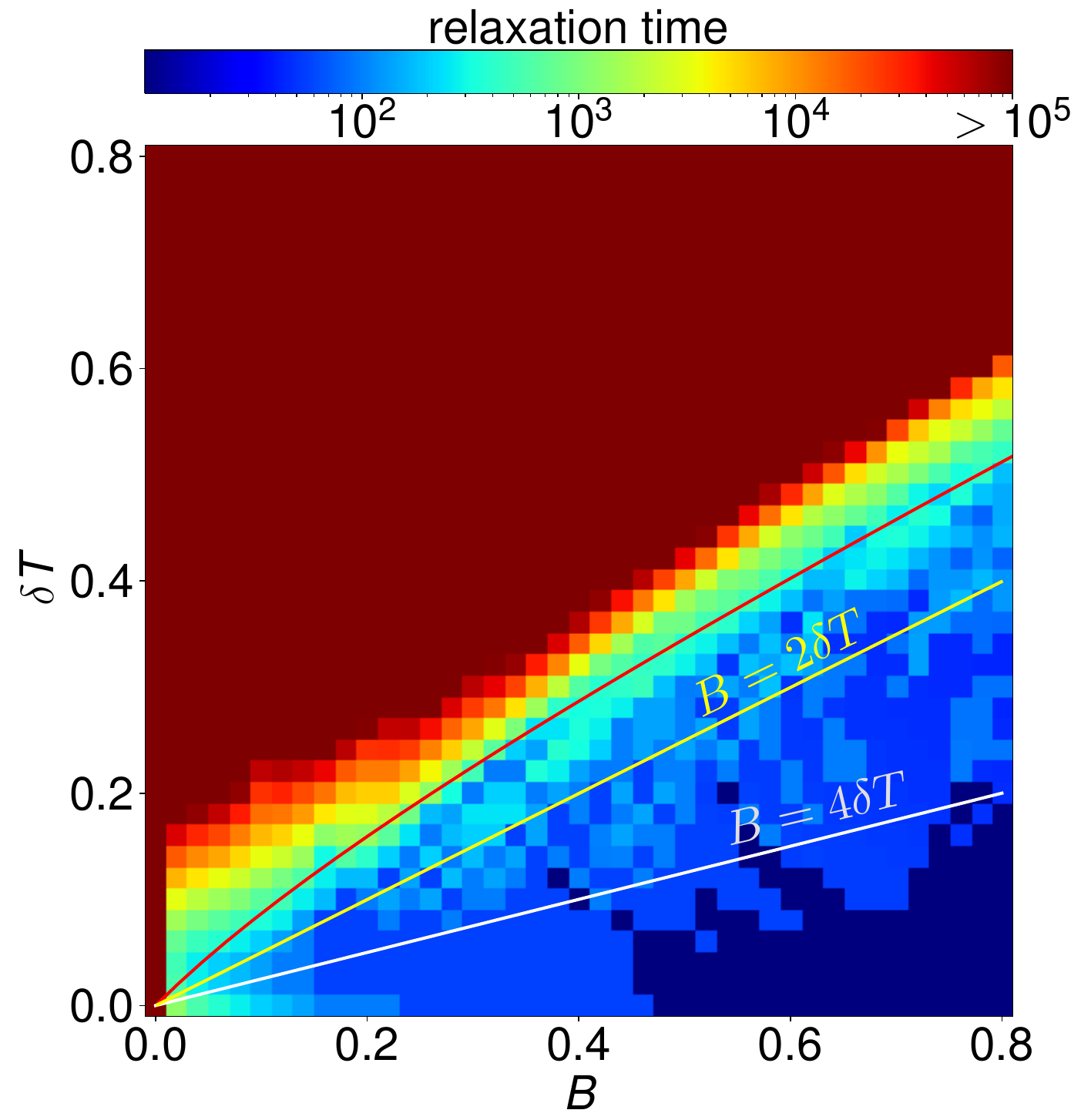}
\caption{
Dynamical phase diagram as in Fig.\ \ref{fig:pd1} but for the topologically nontrivial case with $\delta T>0$ and $L=47$.
$\tau$ (color code): shortest time for which $S_{z}/S>95\%$. 
Dark red: $\tau > 10^{5}$.
White line: $B=4 |\delta T|$. 
Yellow line: $B=2 |\delta T|$. 
Curved red line: see text.
}
\label{fig:pd2}
\end{figure}

The pre-relaxation process in fact takes place in a large parameter range of the dynamical phase diagram.
This is quantified in Fig.\ \ref{fig:pd2}, where the relaxation time $\tau$ is again pragmatically defined as the shortest time for which $S_{z}/S>95\%$. 
This is the same criterion used for the topologically trivial case.
We see that the phase diagram is roughly similar to the one obtained in the absence of the edge state, cf.\ Fig.\ \ref{fig:pd1}.
The parameter range, where (pre-)relaxation is observed, however, considerably extends beyond the $B=4|\delta T|$-line, which, for the topologically trivial case, was obtained from the linear-response approach. 
Furthermore, the relaxation time does no longer increase monotonically with $\delta T$ in general, and we observe certain narrow parameter ranges with comparatively fast pre-relaxation.
We also note that the numerical determination of the pre-relaxation time becomes difficult for very weak $B \lesssim 0.01$.

\begin{figure}[b]
\centering
\includegraphics[width=0.75\columnwidth]{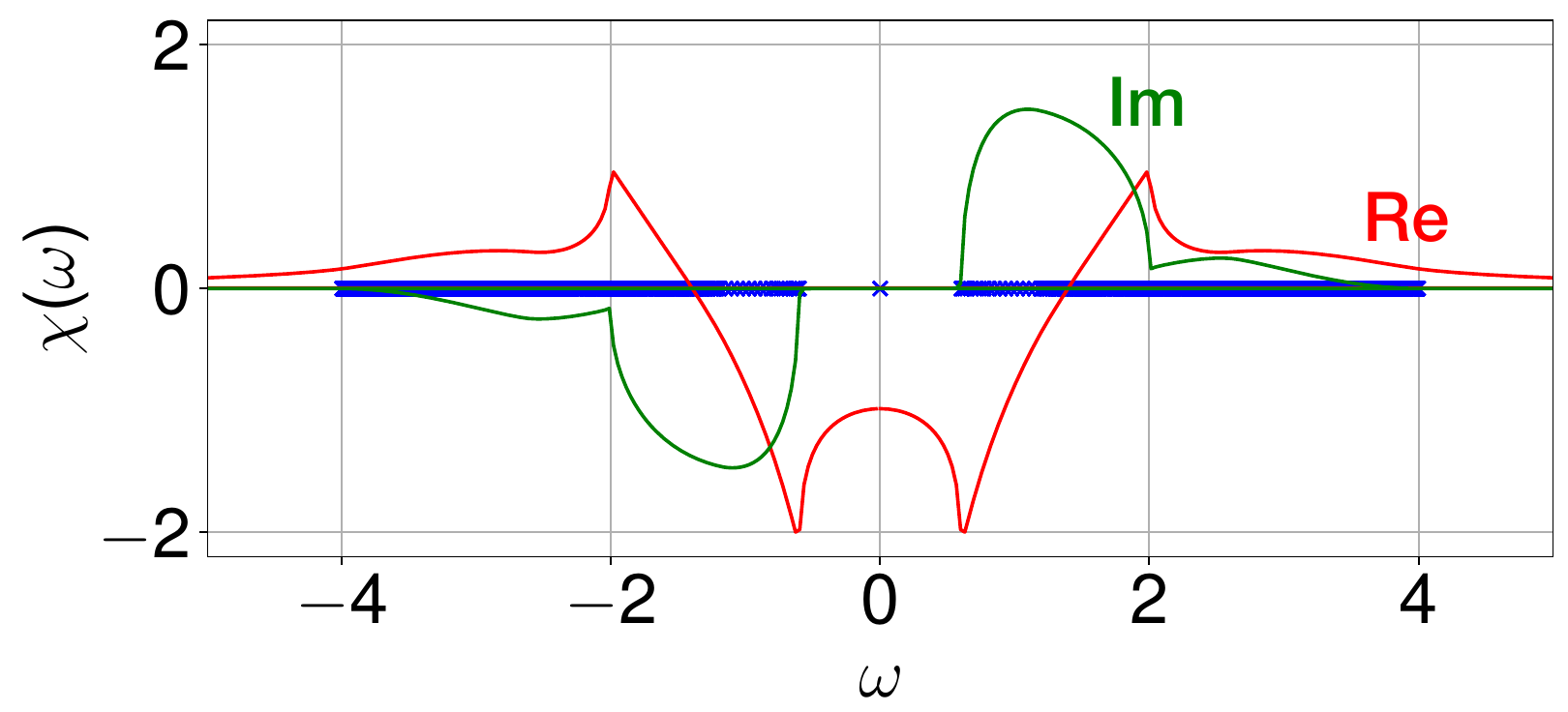}
\caption{
Imaginary and real part of the local retarded spin susceptibility $\chi_{\rm loc}(\omega)$ of the SSH model at site $i_{0}=1$. 
Results for the topologically nontrivial case $\delta T = 0.3$. 
Blue crosses indicate the positions of the poles of $\chi_{\rm loc}(\omega)$. 
Note that there is a pole at $\omega = 0$ lying in the spin gap and that the resulting $\delta$-like peak in the imaginary part of $\chi_{\rm loc}(\omega)$ is not shown. 
At $\delta T = 0.3$, the spin gap is given by $\Delta = 4 \delta T = 1.2$.
}
\label{fig:chi0}
\end{figure}

\begin{figure}[t]
\centering
\includegraphics[width=0.8\columnwidth]{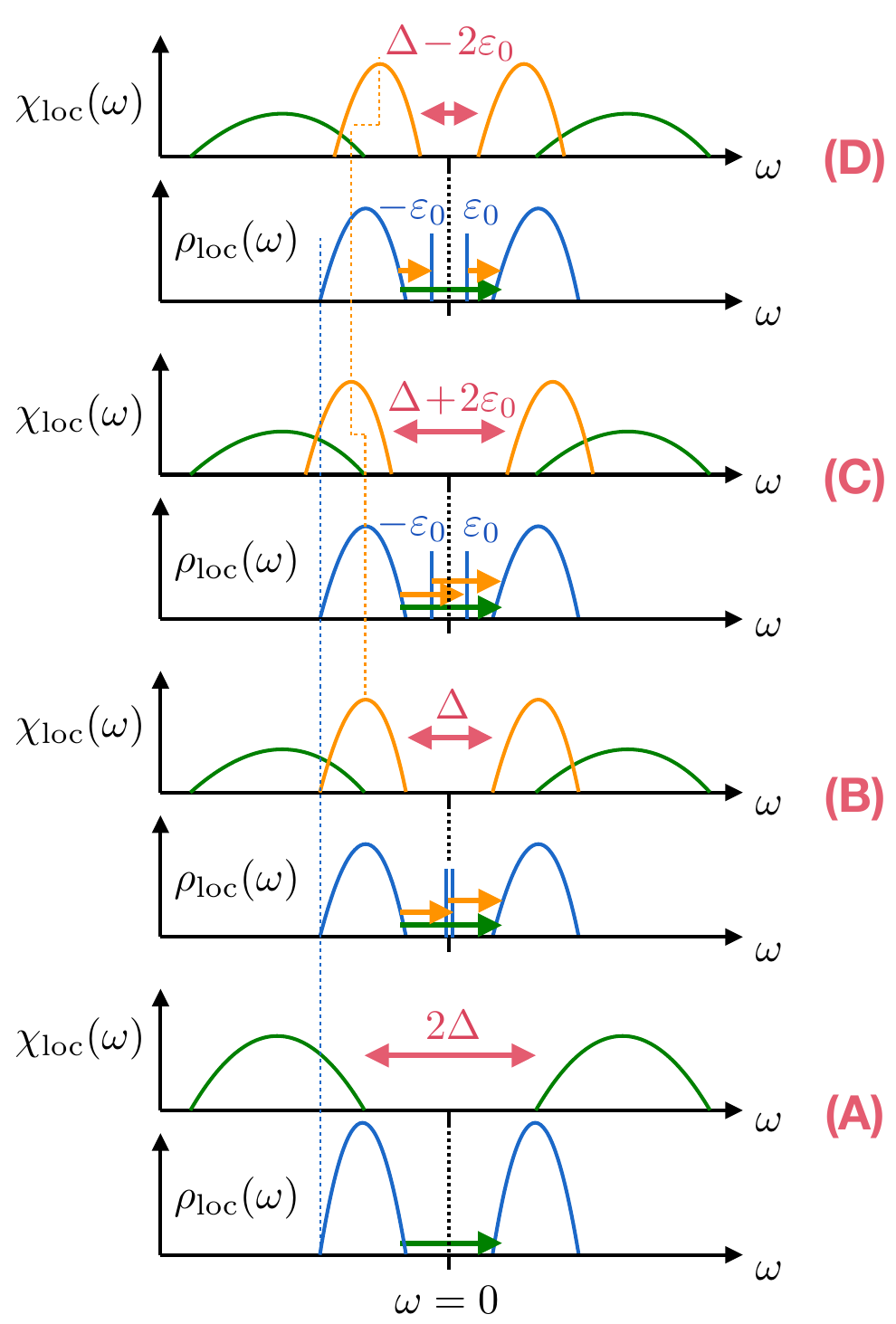}
\caption{
Sketch of the the local density of states $\rho_{\rm loc}(\omega)$ and (absolute value of the) imaginary part of the local spin susceptibility $|\mbox{Im} \chi_{\rm loc}(\omega)|$ at site $i_{0}=1$. 
(A) Topologically trivial case, no edge state. 
(B) Topologically nontrivial case, $J=0$, spin-degenerate edge state.
(C) $J>0$, spin-split edge state.
(D) $J>0$, spin-split edge state, non-equilibrium contributions to $\chi_{\rm loc}(\omega)$.
$\Delta$ is the bulk band gap of the density of states.
Green: particle-hole excitations between occupied/unoccupied extended band states and resulting contributions to $\chi_{\rm loc}(\omega)$.
Yellow: particle-hole excitations involving the localized edge states and resulting contributions to $\chi_{\rm loc}(\omega)$.
Red: resulting spin gap.
}
\label{fig:conv}
\end{figure}

\begin{figure}[t]
\centering
\includegraphics[width=0.75\columnwidth]{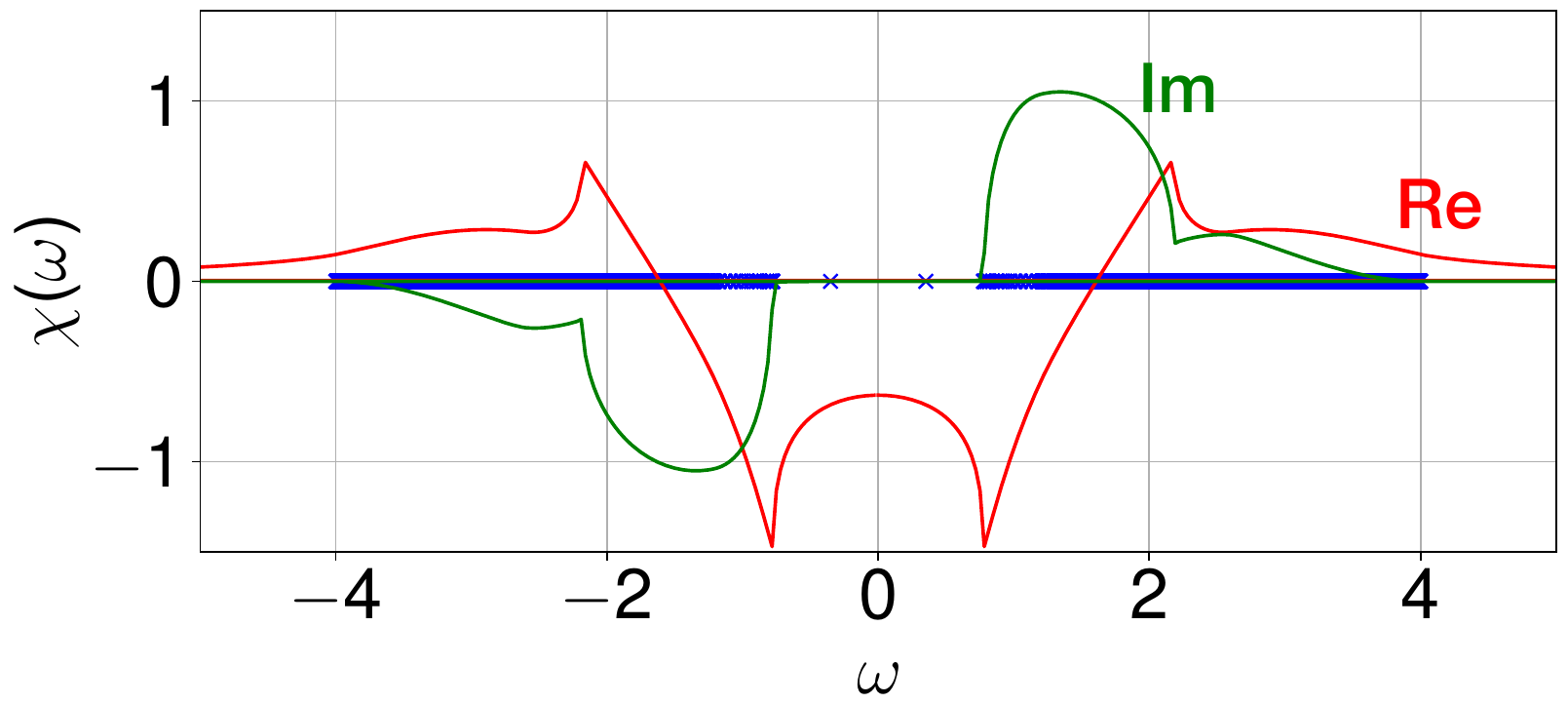}
\caption{
Spin susceptibility $\chi_{\rm loc}^{(xx)}(\omega)$  as in Fig.\ \ref{fig:chi0} but calculated with the effective hopping matrix at $J=1$ for $\bm{S} = S \ff e_x$.
Note that there are two poles at $\omega = \pm 2\varepsilon_{0} \approx \pm 0.351$ lying in the spin gap and that the resulting $\delta$-like peaks in the imaginary part of $\chi_{\rm loc}(\omega)$ are not shown. 
At $\delta T=0.3$, the spin gap is $1.55 \approx \Delta + 2 \varepsilon_{0} > \Delta=4\delta T =1.2$.
}
\label{fig:chi1}
\end{figure}

In an attempt to understand the phase boundary of the (pre-)relaxation regime we consider linear-response theory again.
The $J=0$ retarded local magnetic susceptibility at $i_{0}$=1 is shown in Fig.\ \ref{fig:chi0}. 
However, as compared to the topologically trivial case, Fig.\ \ref{fig:chi}, there are important differences. 
First, the edge state at $\omega=0$ in the LDOS gives rise to a $\delta$-peak in the imaginary part of $\chi_{\rm loc}(\omega)$.
We note that this cannot contribute to the spin damping since, according the condition Eq.\ (\ref{eq:cond}), it corresponds to the static case $B=0$.
Second, however, as the spin-degenerate edge state is filled by exactly one electron, it also mediates additional particle-hole excitations with finite excitation energy.
Single-particle excitations from the highest occupied states of the lower band to the edge state and from the edge state to the lowest unoccupied states of the upper band have an energy of $\omega=2\delta T= \Delta /2$, where $\Delta$ is the bulk band gap.

The susceptibility shown in Fig.\ \ref{fig:chi0} can be understood as the convolution of the occupied with the unoccupied part of the LDOS at $i_{0}=1$.
Hence, due to the presence of the edge state, there are additional contributions to $\chi_{\rm loc}(\omega)$ due to the convolution of the $\delta$-peak, resulting from the edge state, with the occupied and with the unoccupied parts of the local density of extended band states.
This explains the additional broad peaks in $\mbox{Im} \, \chi_{\rm loc}(\omega)$ around $\omega=\pm 1.2$, see Fig.\ \ref{fig:chi0}. 
It also explains that the spin gap in $\chi_{\rm loc}(\omega)$ equals the bulk band gap $\Delta$, i.e., exactly half of the spin gap present in the topologically trivial case. 

Fig.\ \ref{fig:conv} provides a schematic overview. 
Panel (A) of the figure, referring to the topologically trivial case, demonstrates that the LDOS with single-particle excitation gap $\Delta$ results in a spin gap of $2\Delta$.
In the topologically nontrivial case, panel (B), the additional contributions to $\chi_{\rm loc}(\omega)$ resulting from convolutions with the $\delta$-like edge-state peak in the LDOS (yellow peaks), lead to a shrinking of the spin gap by the mentioned factor two.

We infer that, on the basis of standard linear-response theory, spin damping is expected to take place for $B >  \Delta/2  =2\delta T$. 
As is seen in Fig.\ \ref{fig:pd2}, however, this prediction is still by far too restrictive. 
Pre-relaxation is found in a much larger parameter range.
We conclude that linear-response theory cannot adequately explain the dynamical phase diagram and that the full theory is necessary to account for the effects of the edge mode on spin relaxation. 

\subsection{Spin splitting of the edge state}
\label{sec:split}

As it derives from lowest-order perturbation theory in $J$, the linear-response approach is based on the $J=0$ spin susceptibility of the unperturbed SSH model. 
This misses at least two effects. 

First of all, at finite $J$, the exchange coupling induces an internal Zeeman-like spin splitting of the edge state. 
Panel (C) of Fig.\ \ref{fig:conv} indicates two spin-split localized edge states at energies $\omega=\pm \varepsilon_{0}$ (with $\varepsilon_{0}>0$) in the LDOS.
The spin splitting of the edge state should have an impact on the relaxation dynamics. 

In an {\em ad hoc} extension of the linear-response theory, it is tempting to employ the resulting $J>0$ spin susceptibility in the criterion (\ref{eq:cond}) to determine the phase boundary for (pre-)relaxation in the dynamical phase diagram Fig.\ \ref{fig:pd2}, i.e., we replace the bare hopping matrix $\ff T$ by $\ff T^{(\rm eff)}(t=0)$.
We refer to this approach ``renormalized'' linear-response theory.

For weak $J$, the splitting can be calculated approximately by using first-order perturbation theory in $J$, i.e., we treat $\bm{\delta  T} = \ff T^{(\rm eff)}(t=0) - \ff{T}$ as a perturbation of the bare hopping matrix $\ff T$.
Without loss of generality, we can assume that $\bm{S}(0)=S\bm{e}_z$ for the moment being.
This implies that $\bm{\delta T} = \text{diag}(\frac{1}{2}JS,-\frac{1}{2}JS,0,0,0,...)$ is diagonal in the basis of one-particle states $\{ | i , \sigma \rangle \}$. 
We pick the unperturbed eigenvector $| \mbox{edge}, \uparrow \rangle$ of $\ff T$ corresponding to the eigenvalue zero in the spin-$\uparrow$ channel, i.e., the spin-$\uparrow$ edge state.
Then,
\be
  \varepsilon_{0} = \langle  \mbox{edge}, \uparrow  | \bm{\delta T} |  \mbox{edge}, \uparrow \rangle > 0\: . 
\label{eq:pert}  
\ee
For the (spinful) SSH model (with $\sigma=\uparrow$) and in the limit $L\to \infty$, the expansion of the edge state in the one-particle basis states $| i, \sigma \rangle$ is given by \cite{AOP16}
\begin{align}
|  \mbox{edge}, \uparrow \rangle = \sum_{i=0}^{\infty} c_{i} |i , \uparrow \rangle
\: , 
\end{align}
where the coefficients $c_{i} = 0$ for even site index $i$ and $c_{i} = (-T_{1}/T_{2})^{(i-1)/2} c_{1}$ for odd $i$, and with $T_{1}, T_{2}$ given by Eq.\ (\ref{eq:t12}).
The modulus of $c_{1}$ is obtained from the normalization condition $\langle \text{edge},\uparrow | \text{edge},\uparrow \rangle =1$ as $|c_1|^{2} = 1-\left(T_{1}/T_{2} \right)^2$.
With Eq.\ (\ref{eq:pert}) this yields:
\begin{align}
\varepsilon_{0} = \frac{1}{2} J S \left( 1-\left( \frac{T-\delta T}{T+\delta T} \right)^2 \right)
\: .
\label{eq:splitting}
\end{align}
For $J=1$ and $\delta T=0.3$ we get $\varepsilon_{0} \approx 0.178$.

We have numerically computed the $xx$-component of the spin-susceptibility tensor $\chi^{(xx)}_{\rm loc}(\omega)$ for $\ff S(t=0) = S \ff e_{x}$ at $J=1$ and $\delta T=0.3$.
The result is shown in Fig.\ \ref{fig:chi1}.
Opposed to Fig.\ \ref{fig:chi0}, there are {\em two} isolated poles of the susceptibility $\omega = \pm 2\varepsilon_{0}$ resulting from the spin-split edge state. 
The numerical calculation yields $\varepsilon_{0} \approx 0.176$, which is in fact very close to the perturbative result discussed above.

As visualized by panel (C) of Fig.\ \ref{fig:conv}, the spin-splitting of the edge state yields a spin gap $\Delta+2\varepsilon_{0}$ which is {\em larger} than the spin gap $\Delta$ obtained with the  unrenormalized theory.
This is due to the fact that an additional energy $\varepsilon_{0}$ is necessary to make transitions from occupied bulk states to the unoccupied edge state with energy $+\varepsilon_{0}$ possible, and vice versa for transitions from the occupied edge state with energy $-\varepsilon_{0}$ to unoccupied bulk states.
This implies that the parameter region, where spin pre-relaxation is predicted, actually {\em shrinks} when
taking into account that spin excitations are also mediated via the spin-split edge state, and that, therefore, the renormalized linear-response theory does {\em not} lead to an improved description. 

\subsection{Dynamic relaxation mechanism}
\label{sec:dyn}

A second effect missing in the standard linear-response approach is the dynamic occupation of states above and the depopulation of states below the Fermi energy.
This non-equilibrium effect provides an additional important mechanism for spin relaxation. 

Let us assume for a moment that the electron system follows the spin dynamics in a perfectly adiabatic way, i.e., that the state of the electron system at time $t$ is given by the ground-state of the system for the current direction of the classical spin at time $t$. 
The Hamiltonian of the spinful SSH model is invariant under global SU(2) spin rotations. 
This symmetry reduces to a U(1) rotation symmetry around the axis defined by the classical spin in case of a finite coupling $J>0$. 
The local ground-state magnetic moment $\langle \ff s_{i_{0}} \rangle_{t}$ of the electron system at site $i_{0}$ must therefore align to $\ff S(t)$, such that the torque of the electron system on the classical spin [see Eq.\ (\ref{eq:eoms})] vanishes at any instant of time, if the dynamics was perfectly adiabatic.
This would imply that there is no spin damping at all. 
As this is not the case, we can thus safely assume that the dynamics is non-adiabatic. 
In fact, the numerical solution of the equations of motion shows that the local magnetic moment $\langle \ff s_{i_{0}} \rangle_{t}$ is always somewhat {\em behind} the motion of the classical spin $\ff S(t)$. 
This has already been noticed earlier in the context of a topologically trivial model \cite{SP15}.

Generally, a non-adiabatic time evolution does not make a big difference as long as bulk states are concerned, as these are barely affected by the impurity spin anyway. 
The spin-dependent occupation of the edge state, on the other hand, is greatly affected by the retardation effect.
If, at a certain instant of time $t$, the edge state $|\mbox{edge}, \sigma \rangle$ is fully polarized, i.e., $n_{\uparrow}(t)=1$ and $n_{\downarrow}(t)=0$, where $\uparrow, \downarrow$ refers to the momentary quantization axis defined by the direction of $\ff S(t)$, it will stay at least partially polarized with respect to the same direction in space also at a slightly later time $t+\Delta t$.
At time $t+\Delta t$, this represents a nonequilibrium configuration.

This effect leads to a {\em partial} occupation of the upper edge state while the lower edge state becomes less occupied. 
At time $t$ we obtain the momentary spin-split edge states $| \mbox{edge}, \pm \rangle_{t}$ as those eigenstates of the effective hopping matrix $\ff T^{\rm (eff)}(t)$ that are localized close to $i_{0}=1$. 
Their occupation is obtained as $n_{\pm}(t) \equiv {}_{t}\langle \mbox{edge}, \pm | \ff \rho(t) | \mbox{edge}, \pm \rangle_{t}$, where $\ff \rho(t)$ is the one-particle reduced density matrix at time $t$.

\begin{figure}[t]
\centering
\includegraphics[width=0.9\columnwidth]{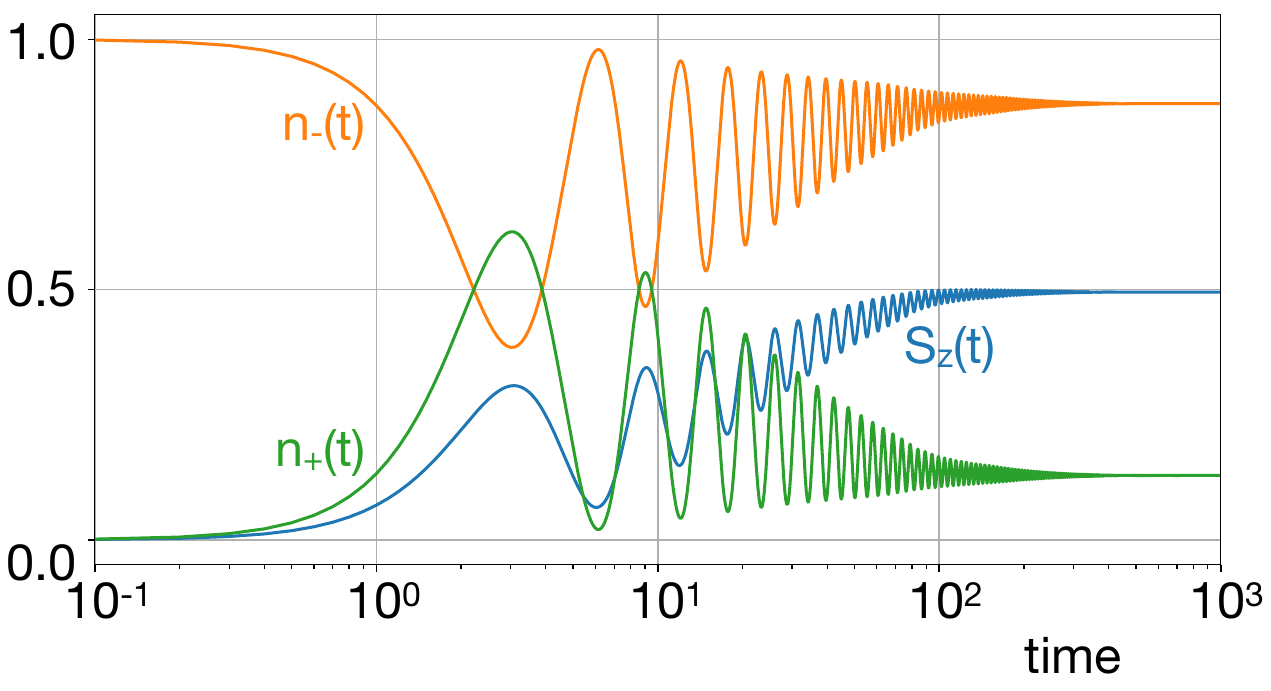}
\caption{
Time evolution of the spin-dependent occupation $n_{\pm}(t)$ of the momentary spin-split edge state, compared to the $z$-component of the classical spin. 
Calculation for $L=47$, $L_B=5$, $\gamma_\text{min}=0.2$, $J=1$, $\delta T=0.3$, and $B=0.75$.
}
\label{occ}
\end{figure}

Fig.\ \ref{occ} displays the time dependence of $n_{\pm}(t)$. 
While at $t=0$, the edge state is fully polarized, there is a partial polarization $n_{+}(t)-n_{-}(t) <1$ for all $t>0$. 
It oscillates with the frequency of the (damped) precessional motion of the classical spin. 
In the long-time limit, a partial polarization survives. 
This again reflects the fact that the spin relaxation is not complete, as has been noted earlier. 

As visualized by panel (D) of Fig.\ \ref{fig:conv}, the non-equilibrium occupation of the spin-split edge state induces additional transitions, namely from the upper edge of the lower bulk band at $-\Delta/2$ to the {\em lower} edge state at $-\epsilon_{0}$, and from the {\em upper} edge state at $\varepsilon_{0}$ to the lower edge of the upper bulk band at $\Delta/2$. 
Within the renormalized linear-response theory this results in a non-equilibrium spin gap of $\Delta-2\varepsilon_{0}$, which in fact is {\em smaller} than the spin gap $\Delta$ obtained with the  unrenormalized theory.
This implies that the parameter region, where spin pre-relaxation is expected, {\em extends}.

Using Eq.\ (\ref{eq:cond}) again, but with the susceptibility of the non-equilibrium state, we can equate $B$ with half of the spin gap and thus get $B = 2\delta T - \varepsilon_0$ as our improved criterion for pre-relaxation. 
This is indicated in the dynamical phase diagram Fig.\ \ref{fig:pd2} by the red solid line. 
We note that this prediction quite convincingly describes the numerical data. 

\subsection{Incomplete spin relaxation}
\label{sec:inc}

An explanation for the incomplete spin relaxation seen in Fig.\ \ref{fig:sd} is still missing and shall be discussed here.
Fig.\ \ref{fig:sd} shows that most of the energy pumped into the system via the sudden flip of the field direction is dissipated to the bulk. 
However, the dissipation process is not complete. 
For times later than the pre-relaxation time scale $\tau$, the impurity spin is close to full alignment with the field $\ff B$ but it is trapped in a state where it steadily precesses around $\ff B$ without any further relaxation. 

We note that the increase of the $z$-component of the impurity spin has stopped at $S_{z} \approx 0.495 < 0.5 = S$ which implies that $\ff S$ encloses a small angle of $\gamma \approx 0.01$ with the field direction.
In this pre-relaxed state the contribution $\ff B \times \ff S$ to the total torque on $\ff S$ can be neglected safely. 
The total torque is then dominated by the torque resulting from the local moment at $i_{0}=1$, i.e., by $J \langle \bm{s}_{i_0} \rangle_t \times \bm{S}(t)$.

As can be read off from Fig.\ \ref{occ}, there is a contribution of about $(0.75 - 0.25)/2 = 0.25$ to the size of the local moment $|\langle \bm{s}_{i_0} \rangle_t|$ due to the polarization of the edge state. 
We assume that this is the dominant contribution to the moment $|\langle \bm{s}_{i_0} \rangle_t|$ at site $i_{0}$. 
This implies that, at $J=1$, the precession frequency is $\omega_{\rm p} \approx 0.25$. 
In fact, we can read off $\omega = 0.285$ from the numerical calculation, see Fig.\ \ref{fig:sd}, supporting this assumption.

Varying parameters, we first of all find that $\omega_{\rm p}$ is almost independent of $B$, which again reflects that the field contribution to the torque is negligible.
Furthermore, the edge-state polarization can be increased when increasing $\delta T$ since this 
increases the spin splitting of the edge state according to Eq.\ (\ref{eq:splitting}), and thus $|\langle \bm{s}_{i_0} \rangle_t|$ and therewith $\omega_{\rm p}$ should increase with increasing $\delta T$. 
This is verified numerically as well. 
As a function of $J$, the precession frequency $\omega_{\rm p}$ increases approximately linearly with $J$ in the range $0 < J \lesssim 1$. 
Also this trend is easily explained since, apart from the retardation effect discussed in Sec.\ \ref{sec:dyn}, the edge state is strongly polarized at $\delta T = 0.3$, so that the $J$ dependence of $\omega_{\rm p}$ is almost exclusively due to bare coupling constant itself.
All in all these considerations explain the transition from a precessional motion with Larmor frequency $\omega_{\rm p} \approx B$ at early times to a frequency $\omega_{\rm p} \approx J |\langle \bm{s}_{i_0} \rangle_t|$ in the pre-relaxed state at late times.

They also explain that the relaxation process must stop at some point.
Recall that the basic argument explaining the dynamic phase boundary for spin relaxation fundamentally builds on the assumption that the impurity spin dynamics can be described, to a good approximation, by a precessional motion with a well-defined precession frequency. 
At early times this is in fact given by the Larmor frequency $\omega_{\rm p}=B$.
At late times, however, the field contribution is negligible, so that we have to apply our basic criterion for spin relaxation with $B$ replaced by the actual precession frequency $\omega_{\rm p} \approx J |\langle \bm{s}_{i_0} \rangle_t|$ in the pre-relaxed state.
This yields $\omega_{\rm p} > 4\delta T$ as a condition for spin damping, and with $\omega_{\rm p} = 0.285$ at late times and $4\delta T = 1.2$, this condition is clearly violated.

We conclude that the observed spin pre-relaxation scenario first of all requires that the field $B$ is sufficiently strong, compared to the band gap, to get initial spin relaxation, and that, at early times, the contribution of $\ff B \times \ff S$ to the total torque on $\ff S$ is dominating. 
As the initial relaxation proceeds, the further and further alignment of $\ff S$ and $\ff B$ just implies that this contribution diminishes. 
In the course of time there may be a gradual transition to a regime where the electronic contribution $J \langle \bm{s}_{i_0} \rangle \times \bm{S}$ to the total torque on $\ff S$ becomes dominant. 
In such a case, even if $B$ would be strong enough to feature relaxation, only the relation between the precession frequency at late times $J |\langle \bm{s}_{i_0} \rangle|$ and the gap size decides whether or not there is full relaxation or whether the system dynamics is finally trapped in a pre-relaxed steady state.

A dominating electronic contribution is not so much favored by a strong coupling constant $J$ since this suppresses the retardation effect and enforces alignment of $\langle \bm{s}_{i_0} \rangle$ and $\ff S$. 
Strong $J$ thus rather leads to a small electronic contribution to the total torque. 
Much more important is a strong polarizability of the local electronic moment $\langle \bm{s}_{i_0} \rangle$ at site $i_{0}$ due to the classical spin $\ff S$ at intermediate $J$.
The presence of the topological edge state very much enhances this polarizability as it is precisely half-filled.
However, the polarizability also crucially depends on the spatial extension of the edge state and is at a maximum for an edge state that is completely localized at $i_{0}$.
Hence, one may argue that a late transition to a pre-relaxed steady state is more characteristic for a one-dimensional system opposed, e.g., to a topological surface state at the boundary of a two-dimensional topological insulator which has a finite one-dimensional dispersion and extends along a one-dimensional boundary, such that the local polarization at $i_{0}$ will be much weaker. 

\subsection{Direct comparison with linear-response theory}
\label{lr}

\begin{figure}[t]
\centering
\includegraphics[width=0.8\columnwidth]{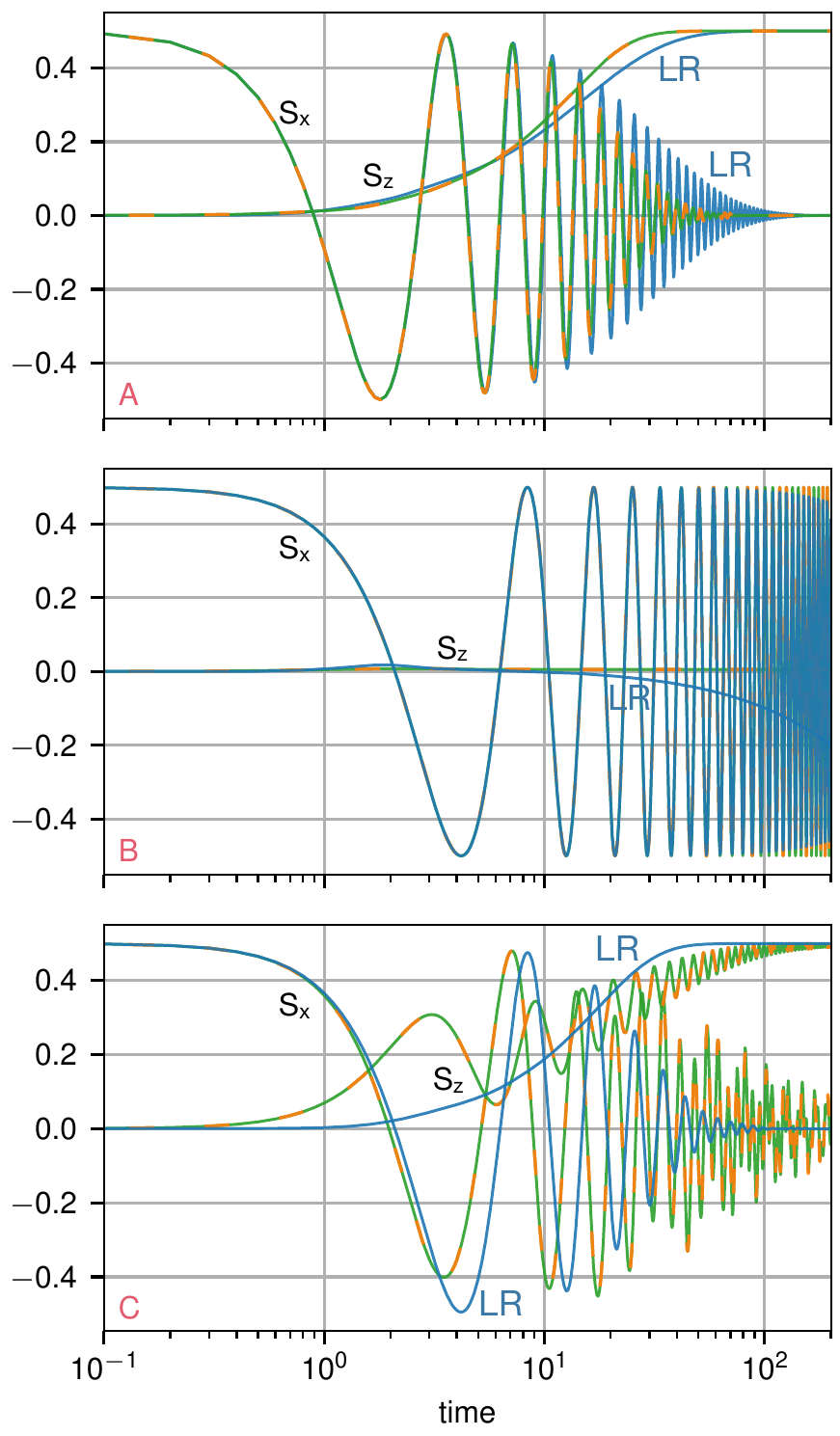}
\caption{
Real-time dynamics of the impurity spin after a sudden flip of the local field from $x$- to $z$-direction. 
Panels A, B: $\delta T = - 0.3$ (topologically trivial) with $B=1.75 > 4 \delta T$ (A) and $B=0.75 < 4 \delta T$ (B). 
Panel C: $\delta T = + 0.3$ (non trivial), $B=0.75 < 4 \delta T$.
Results as obtained from three different approaches: 
Blue lines: linear-response theory (LR) for $L=500$ (panels A,B) and $L=501$ (panel C).
Green lines: spin-dynamics theory for $L=200$ (A,B) and $L=201$ (C) with open boundaries.
Orange lines: spin-dynamics theory for $L=46$ (A,B) and $L=47$ (C) and absorbing boundaries (with $L_{B}=20$, $\gamma_\text{min}=0.05$ in A, and $L_{B}=5$, $\gamma_\text{min}=0.2$ in B,C).
In all cases $J=1$. Note the logarithmic time scale.
}
\label{fig:comp}
\end{figure}

Our arguments explaining the absence or the occurrence of spin relaxation are build on the framework of linear-response theory, and we have already seen that various inconsistencies show up when naively applying this approach and that refined considerations are necessary for the correct physical picture.
Therefore, a {\em direct} comparison of linear-response theory with the full spin-dynamics theory should be instructive. 
Fig.\ \ref{fig:comp} displays the predictions of linear-response (LR) and full theory for the impurity-spin dynamics in various parameter regimes. 
Panel A gives an example for the topologically trivial case ($\delta T = - 0.3$), similar to the upper panel in Fig.\ \ref{fig:relax}.
The field $B=1.75$ is larger than the gap $4|\delta T|=1.2$ and, therefore, we expect complete relaxation of the impurity spin.
This is in fact found by LR theory when numerically solving the integro-differential equation (\ref{eq:ide}), see blue lines in A.
However, when compared with the full theory, either for a larger system with $L=200$ and open boundaries, or for a smaller system with $L=46$ and an absorbing boundary (on the edge opposite to the impurity spin), see green and orange lines, respectively, strong differences are visible. 
Most notably, the LR approach significantly overestimates the spin-relaxation time, by about a factor of two.
On the other hand, there is no visible discrepancy between the calculations with open and with absorbing boundaries.

In panel B the field $B=0.75$ is smaller than the gap $4|\delta T|=1.2$, and hence relaxation is not expected  to take place (cf.\ the lower panel of Fig.\ \ref{fig:relax}).
Full spin-dynamics theory with open and with absorbing boundary do not differ visibly. 
In both cases, the impurity spin shows an undamped precessional motion on the time interval displayed. 
Calculations with absorbing boundaries can be performed for much longer time scales and in fact do not show any relaxation effect at all (not shown). 
The numerical evaluation of linear-response theory is much more involved, such that we are basically restricted to the shorter time scale displayed in the figure. 
We note that already at early times, $t \approx 10$, the impurity spin starts to ``relax'', i.e., it develops a finite and even {\em negative} $z$-component. 
This is not only at variance with the full spin-dynamics theory but also unphysical. 

Finally, in panel C we compare the three approaches for the topologically nontrivial case.
The real-time dynamics is more complicated in this case. 
The calculations performed with open and with absorbing boundary conditions match perfectly and show the pre-relaxation behavior as discussed above (cf.\ Fig.\ \ref{fig:sd}).
Linear-response theory, on the other hand, still predicts {\em complete} spin relaxation, and the LR results look more similar to those obtained for the topologically trivial case in panel A.
We conclude that linear-response theory is not able to reproduce the correct physics quantitatively but also qualitatively. 
A deeper analysis of the causes of the difficulties of LR theory does not seem worthwhile, since spin-dynamics calculations with absorbing edges are numerically much easier and faster. 
Still, the criterion for spin relaxation, Eq.\ (\ref{eq:cond}), derived within the linear-response theory, has proven as a rough but valuable guide for the understanding of the dynamical relaxation phase diagrams. 

\section{Conclusions}
\label{con}

The spinful SSH model with a classical impurity spin exchange-coupled to the local magnetic moment of the electron system at one of the edge sites probably represents the simplest system which allows us to study the effect of a topological edge mode on the relaxation of the impurity spin.
Using an absorbing boundary at the opposite edge, we have demonstrated that the coupled dynamics of impurity spin and electron system can be treated numerically exactly. 
We have used this approach to study the spin relaxation time. 

The most obvious effect is the presence or absence of spin relaxation in certain parameter regimes. 
One can in fact construct a dynamical phase diagram with a rather sharp phase boundary. 
Comparing the phase diagrams for topologically trivial and nontrivial bulk electronic structures, reveals significant differences which must be attributed to the absence of presence of the topologically protected edge state, respectively.

For a qualitative understanding of the various dependencies on the relevant model parameters, however, 
a simplified framework is needed.
We have argued that this is basically provided by the linear-response theory, i.e., by lowerst-order perturbation theory in the coupling strength $J$.

Admittedly, the linear-response approach can dramatically fail in predicting the microscopic real-time dynamics, as has been demonstrated by direct comparison with the results of the full spin-dynamics theory. 
Furthermore, its numerical evaluation requires the solution of an integro-differential equation and is thus much more costly compared to the full theory which must take the time-dependent electronic degrees of freedom into account explicitly but, thanks to the absorbing boundary, can be restricted to small system sizes.
On the other hand, the linear-response theory provides an extremely handy and physically appealing criterion for spin relaxation based on the gap size of magnetic excitations. 

We have shown that the naive application of this condition for spin relaxation works in the topologically trivial case but must be corrected in the nontrivial case. 
The most important mechanisms affecting the spin relaxation in the nontrivial case are the following: 
The presence of the edge state mediates additional spin-exchange processes and thus leads to a strong extension of the parameter regime where relaxation is possible. 
A renormalized linear-response theory is necessary to account for the effects of the spin splitting of the edge state. 
In a static picture, however, the spin splitting tends to suppress spin relaxation since the energy of the lowest-energy spin excitations involving the spin-split edge mode increases. 
We find that this is not the correct point of view and that one must apply a dynamical picture beyond the adiabatic approximation. 
Namely, the time evolution of the electronic structure and of the edge state in particular is considerably retarded and does not instantaneously follow the motion of the impurity spin. 
This causes a partial occupation of the upper edge state with energy above the Fermi energy, and vice versa some depopulation of the lower state of the spin-split Kramers pair, which in turn opens another lowest-energy channel for relaxation and in fact finally correctly describes the additional extension of the parameter regime for spin relaxation.

At late times where spin relaxation is almost completed, the magnetic-field term can be disregarded and the spin torque due to the local magnetic moment of the electron system may dominate. 
It then leads to a precessional motion which can be slowed down decisively due to the mentioned dynamical partial polarization of the edge mode such that the criterion for spin relaxation gets violated and the relaxation process comes to a halt. 
The resulting pre-relaxation or incomplete spin-relaxation phenomenon is frequently found in the numerical solution of the full set of equations of motion. 
However, it might represent an effect that is characteristic for one-dimensional systems only. 

The present study has paved the way for investigations of the dynamics of individual spins or of spin arrays on two-dimensional Chern or Z$_{2}$ insulators, where new phenomena but also additional complications are expected. 
Some important differences will be due to the delocalization of the topological edge state along an extended one-dimensional edge such that its spin polarization will diminish, or additional features brought in by spin-momentum locking. 
Work along these lines is in progress.

\acknowledgments
This work was funded by the Deutsche Forschungsgemeinschaft (DFG, German Research
Foundation) through the Cluster of Excellence ``Advanced Imaging of Matter'' - EXC 2056 - project ID 390715994, and through the Collaborative Research Center ``Light-Induced Dynamics and Control of Correlated Quantum Systems'' - SFB 925 - project ID 170620586 (project B5).


%

\end{document}